\documentclass[a4paper]{JHEP3}
\usepackage[latin1]{inputenc}
\usepackage{bbm}
\usepackage{bm}
\usepackage{graphicx}
\usepackage{epsfig}
\usepackage{subfigure}
\usepackage{latexsym}
\usepackage{amsmath}
\usepackage{amsfonts}
\usepackage{amssymb}
\usepackage{color}

\def\tr{\mathrm{Tr}}

\newcommand{\nf}{$\mathcal{N}=4$ }
\newcommand{\beq}{\begin{equation}}
\newcommand{\eeq}{\end{equation}}

\newcommand{\be}{\begin{eqnarray}}
\newcommand{\ee}{\end{eqnarray}}

\definecolor{rossoCP3}{cmyk}{0,.88,.77,.40}
\usepackage{fancyhdr}
\title{\textcolor{rossoCP3} {\nf Extended MSSM}}
\author{Matti Antola\footnote{matti.antola@helsinki.fi}\\
Department of Physics and Helsinki Institute of Physics, 
P.O.Box 64, FI-000140, University of Helsinki, Finland}
\author{Stefano Di Chiara\footnote{dichiara@cp3.sdu.dk}\\
{ CP}$^{ \bf 3}${-Origins}, 
Campusvej 55, DK-5230 Odense M, Denmark}
\author{Francesco Sannino\footnote{sannino@cp3.sdu.dk}\\
{ CP}$^{ \bf 3}${-Origins}, 
Campusvej 55, DK-5230 Odense M, Denmark}
\author{Kimmo Tuominen \footnote{kimmo.i.tuominen@jyu.fi}\\
Department of Physics,
P.O.Box 35, FI-40014, University of Jyv\"askyl\"a, Finland, and\\
Helsinki Institute of Physics, 
P.O.Box 64, FI-000140, University of Helsinki, Finland}

\abstract
{We investigate a perturbative \nf sector coupled to the MSSM and show that it allows for a stable vacuum correctly breaking the electroweak symmetry. The particle spectrum of the MSSM is  enrichened by several new particles stemming out from the new \nf sector of the theory, and a new lepton doublet required to cancel global and gauge anomalies of the theory.  Even if the conformal invariance of the \nf sector is explicitly broken, a nontrivial UV behavior of the coupling constants is possible: by studying the renormalization group equations at two loops we find that the Yukawa couplings of the heavy fermionic states flow to a common fixed point at a scale of a few TeVs. The parameter space of the new theory is reduced imposing naturalness of the couplings and soft supersymmetry breaking masses, perturbativity of the model at the EW scale as well as phenomenological constraints. Our preliminary results on the spectrum of the theory suggest that the LHC can rule out a significant portion of the parameter space of this model.
\\
{\footnotesize  \it Preprint: CP$^3$-Origins-2010-37}}

\maketitle

\begin{document}

\section{Introduction}
\label{introduction}

In this work we investigate the phenomenological viability of a perturbative \nf  Super Yang--Mills (4SYM) sector coupled to the Minimal Supersymmetric Standard Model (MSSM).

The historical reason for arriving to this model was to provide a UV completion for the Minimal Walking Technicolor (MWT) model. The field content of MWT is constituted by two flavors of fermions and one gluon all in the adjoint representation of the new gauge group $SU(2)_{N4}$. The electroweak (EW) symmetry is broken by a technifermion condensate. The need to go beyond this pure technicolor theory arises from the necessity to generate SM fermion masses. One possibility is to reintroduce new bosons ({\it bosonic technicolor}) \cite{Simmons:1988fu,Kagan:1991gh,Carone:1992rh,Carone:1994mx,Antola:2009wq} able to give masses to the SM fermions using standard Yukawa interactions. Supersymmetric technicolor has been considered \cite{Dine:1981za,Dobrescu:1995gz} as a way to naturalize bosonic technicolor.

Interestingly, in \cite{Antola:2010nt} we realized that the fermions and gluons of Minimal Walking Technicolor fit perfectly in an \nf supermultiplet, provided that we also include three scalar superpartners. In fact the $SU(4)$ global symmetry of MWT is simply the $SU(4)_R$ $R$-symmetry of the 4SYM theory. This is the global quantum symmetry that does not commute with the supersymmetry transformations. We found that supersymmetrizing MWT in this way leads to an approximate \nf supersymmetry of the technicolor sector, broken only by EW gauge and Yukawa interactions. Due to the approximate \nf symmetry, the beta function of the supersymmetrized technicolor gauge coupling vanishes at one loop. We called this model Minimal Supersymmetric Conformal Technicolor (MSCT).

In this work we abandon the technicolor perspective by considering a perturbative \nf sector at the EW scale. This model constitutes an independent extension of MSSM featuring maximal supersymmetry in four dimensions. The \nf symmetry is broken to $\mathcal{N}=1$ when embedding the MSSM EW sector inside the SU$(4)$ R-symmetry.

In the next section we introduce the particle content and interactions of the model. We then determine the spectrum of the theory. The lightest CP-even and -odd Higgses, massless at tree level, will acquire mass at one loop. We briefly discuss the phenomenologically viability of the spectrum (section \ref{PhV}), concentrating especially on the light scalar states, in light of current and upcoming data from the Large Hadron Collider (LHC). The conformal invariance of the \nf sector is explicitly broken by EW gauging, Yukawa couplings to the MSSM, and soft SUSY breaking terms. However, we discover a possible nontrivial UV behavior of the coupling constants: by studying the renormalization group equations at two loops we find that the Yukawa couplings of the heavy fermionic up-type states flow to a common fixed point at a scale of a few TeVs, while the remaining ones go to zero with increasing energy. The \nf coupling, whose one loop beta function is zero, at two loops instead monotonically decreases for increasing energy.

\section{The Model}
\label{model}

The particle content of the model is given concisely in Table \ref{MSCTsuperfields}, in terms of ${\mathcal{N}}=1$ superfields. We denote the superfields of 4SYM sector as follows:
\beq
\left(\tilde{U}_L,\ U_L\right)\in \Phi_1,\quad \left(\tilde{D}_L,\ D_L\right)\in \Phi_2,\quad   
\left(\tilde{\bar{U}}_R,\ \bar{U}_R\right)\in \Phi_3,\quad \left(G,\ \bar{D}_R\right)\in V,  
\label{superq}
\eeq
where we used a tilde to label the scalar superpartners. We indicated with $\Phi_i$, $i=1,2,3$ the three chiral superfields of 4SYM and with $V$ the vector superfield. To accomodate for the Witten anomaly, induced by the introduction of the \nf sector with respect to the weak interactions, we introduce a new lepton doublet constituded by superfields $N$ and $E$:
\beq
\left(\tilde{N}_L,\ N_L\right)\in \Lambda_1,\quad \left(\tilde{E}_L,\ E_L\right)\in \Lambda_2,\quad   
\left(\tilde{\bar{N}}_R,\ \bar{N}_R\right)\in N,\quad\left(\tilde{\bar{E}}_R,\ \bar{E}_R\right)\in E.
\label{superl}
\eeq
\begin{table}[h]
\centering
\begin{tabular}{c|c|c|c|c}
\noalign{\vskip\doublerulesep}
Superfield & SU$(2)_{N4}$ & SU$(3)_{\text{c}}$ & SU$(2)_{\text{L}}$ & U$(1)_{\text{Y}}$\tabularnewline[\doublerulesep]
\hline
\noalign{\vskip\doublerulesep}
$\left(\begin{array}{c}\Phi_{1}\\ \Phi_{2}\end{array}\right)$ & Adj & $1$ & $\square$ & 1/2\tabularnewline[\doublerulesep]
\noalign{\vskip\doublerulesep}
$\Phi_{3}$ & Adj & 1 & 1 & -1\tabularnewline[\doublerulesep]
\noalign{\vskip\doublerulesep}
$V$ & Adj & 1 & 1 & 0\tabularnewline[\doublerulesep]
\noalign{\vskip\doublerulesep}
$\left(\begin{array}{c}\Lambda_{1}\\ \Lambda_2\end{array}\right)$ & 1 & 1 & $\square$ & -3/2\tabularnewline[\doublerulesep]
\noalign{\vskip\doublerulesep}
$N$ & 1 & 1 & 1 & 1\tabularnewline[\doublerulesep]
\noalign{\vskip\doublerulesep}
$E$ & 1 & 1 & 1 & 2\tabularnewline[\doublerulesep]
\noalign{\vskip\doublerulesep}
$H$ & 1 & 1 & $\square$ & 1/2\tabularnewline[\doublerulesep]
\noalign{\vskip\doublerulesep}
$H'$ & 1 & 1 & $\square$ & -1/2\tabularnewline[\doublerulesep]
\end{tabular}\caption{The particle content of the model in terms of the ${\cal N}=1$ superfields}
\label{MSCTsuperfields}
\end{table}

The renormalizable lepton and baryon number\footnote{We assume all the superfields in Table \ref{MSCTsuperfields} to have both lepton and baryon numbers equal to zero.} conserving superpotential for the model is 
\beq 
P=P_{MSSM}+P_{N4}, \label{spot} 
\eeq 
where $P_{MSSM}$ is the minimal supersymmetric standard model (MSSM) superpotential, and
\beq 
P_{N4}=-\frac{g_{N4}}{3\sqrt{2}} \epsilon_{ijk} \epsilon^{abc} \Phi^a_i \Phi^b_j \Phi^c_k+y_U \epsilon_{ij}\Phi^a_i H_j\Phi^a_3+y_N \epsilon_{ij}\Lambda_i H_j N+y_E \epsilon_{ij}\Lambda_i H^{\prime}_j E+y_R E \Phi_3^a \Phi_3^a. 
\label{spmwt} 
\eeq
In the last equation $\Phi_{i}^{a}=Q_{i}^{a}$, $i=1,2$, with $a$ the 4SYM index. The first coefficient, $g_{N4}$, is the gauge coupling of the 4SYM theory. The presence of the \nf gauge coupling here is a remnant of the \nf nature of this operator. We have explicitly verified in Appendix \ref{rge} that starting instead from a generic value $y_{N4}$ at some UV scale, one recovers $y_{N4}=g_{N4}$ at the EW scale.

The Lagrangian of the model reads
\beq
{\cal L}={\cal L}_{MSSM}+{\cal L}_{N4} \ ,
\label{smwt}
\eeq
where ${\cal L}_{N4}$, by following the notation of Wess and Bagger \cite{Wess:1992cp}, is:
\beq
{\cal L}_{N4}=\frac{1}{2} \tr \left(W^{\alpha} W_{\alpha}|_{\theta\theta}+\bar{W}_{\dot{\alpha}} \bar{W}^{\dot{\alpha}}|_{\bar{\theta}\bar{\theta}}\right)+\Phi_f^{\dagger}\exp \left( 2 g_X V_X \right) \Phi_f|_{\theta\theta\bar{\theta}\bar{\theta}}+\left(P_{N4}|_{\theta\theta}+h.c.\right),
\label{tecL}
\eeq
In the last equation 
\beq W_\alpha=-\frac{1}{4 g}\bar{D}\bar{D}\exp\left(-2 g V\right)D_\alpha\exp\left(2 g V\right),\ \ V=V^a T^a_A,\ \  \left(T^a_A\right)^{bc}=-i f^{abc}, 
\eeq
and
\beq
\Phi_f=Q,\Phi_3,\Lambda,N,E;\quad X=N4, L, Y\ .
\eeq
The product $g_X V_X$ includes the gauge charge of the superfield on which it acts. The charge is $Y$ for $U(1)_Y$, and is 1 (0) for a multiplet (singlet) of a generic group $SU(N)$. The 4SYM vector superfield $V_{N4}$ is $V$ defined in Eq.(\ref{superq}). The remaining vector superfields are those of the MSSM \cite{Martin:1997ns}, while the superpotential $P_{N4}$ is given in Eq.(\ref{spmwt}). For the benefit of the reader, the MSSM extension, ${\cal L}_{N4}$, is in Appendix \ref{appxA}. The full MSSM Lagrangian ${\cal L}_{MSSM}$ can be found in \cite{Martin:1997ns} and references therein.

Of course, any supersymmetry must break, and following the MSSM literature we do so by adding SUSY breaking soft terms. These are given explicitly in Eq. \eqref{Lsof}.

\section{Vacua and Stability Conditions}
\label{VacM}

To derive the spectrum of the theory, we first determine the model's ground state. We allow for a nonzero vacuum expectation value (vev) for each of the electromagnetically neutral scalars, which are $\tilde{D}_L$, $\tilde{H}_0$ and $\tilde{H}_0^\prime$. Without loss of generality, we choose the vacuum expectation value of the $\tilde{D}_L$ scalar to be aligned in the third direction of the $SU(2)_{N4}$ gauge space, and hence the vacuum expectation values (vevs) are written as
\beq
\left<\tilde{D}_L^3\right>=\frac{v_{N4}}{\sqrt{2}},\quad  \left<\tilde{H}_0\right>=s_\beta \frac{v_H}{\sqrt{2}},\quad  \left<\tilde{H}_0^\prime\right>=c_\beta \frac{v_H}{\sqrt{2}},
\label{veveq}
\eeq
where $s_\beta=\sin\beta$, $c_\beta=\cos\beta$, and all vevs are chosen to be real. We indicated the scalar component of each Higgs weak doublet superfield with a tilde. From these we find that the gauge group breaking follows the pattern $SU(2)_{N4}\times SU(2)_L\times U(1)_Y\rightarrow U(1)_{N4}\times U(1)_{EM}$. The second $U(1)$ on the right corresponds to the ordinary electromagnetic (EM) charge.

The neutral scalar potential is obtained from the $D,\ F$, and $soft$ terms of the Lagrangian given in Appendix \ref{appxA}, and from the corresponding MSSM scalar potential. The resulting potential is:
\be
\label{vin}
V_{in}&=&M_Q^2 |\tilde{D}_L^3|^2+\left(m_u^2+|\mu|^2\right) |\tilde{H}_0|^2+\left(m_d^2+|\mu|^2\right) |\tilde{H}_0^\prime|^2-\left(b\tilde{H}_0\tilde{H}_0^\prime+c.c.\right)\nonumber\\
&+&\frac{1}{8}\left(g_L^2+g_Y^2\right)\left(|\tilde{D}_L^3|^2-|\tilde{H}_0^\prime|^2+|\tilde{H}_0|^2\right)^2.
\ee
The terms depending on the phase of the different fields are the $b$ term and its conjugate. As in the MSSM, the invariance under $U(1)_Y$ symmetry, together with the fact that $\tilde{H}$ and $\tilde{H}^\prime$ have opposite hypercharges, allows to redefine their vevs and $b$ parameter to be real.

The quartic terms in this potential cancel when $|\tilde{D}_L^3|^2=|\tilde{H}_0^\prime|^2-|\tilde{H}_0|^2$. To make the potential bound from below we impose the Hessian of $V_{in}$ to be semi-definite positive along this $D$ flat plane, which gives the conditions:
\beq
 \left(m_u^2+|\mu|^2-M_Q^2\right)\left(m_d^2+|\mu|^2+M^2_Q\right)\geqslant b^2,\ 2|\mu|^2+m_u^2+m_d^2\geqslant 0 \ .
\label{cond1}
\eeq
The extremum is defined by
\beq
\partial_{\tilde{D}_L^3} V_{in}|_{\phi=<\phi>}=0,\quad \partial_{\tilde{H}_0} V_{in}|_{\phi=<\phi>}=0,\quad \partial_{\tilde{H}_0^\prime} V_{in}|_{\phi=<\phi>}=0\ .
\eeq
These equations can be used to express the soft SUSY breaking parameters according to:
\be
M_Q^2&=&-\frac{1}{8}\left(g^2_L+g^2_Y\right) \left(v_{N4}^2-c_{2\beta} v_H^2\right),\label{v1}\\
  m_u^2&=&-\frac{1}{8}\left(g^2_L+g^2_Y\right) \left(v_{N4}^2-c_{2\beta} v_H^2\right)-|\mu|^2+b\ t^{-1}_\beta,\label{v2}\\
     m_d^2&=&\frac{1}{8}\left(g^2_L+g^2_Y\right) \left(v_{N4}^2-c_{2\beta} v_H^2\right)-|\mu|^2+b\ t_\beta, \label{v3}
\ee
where $t_\beta=\tan\beta$. Requiring the potential to be unstable at the origin along the $|\tilde{H}^\prime_0| $, $|\tilde{H}_0|$ and $|\tilde{D}^3_L|$ directions gives
\beq
M_Q^2<0,\qquad \left(m_u^2+|\mu|^2\right)\left(m_d^2+|\mu|^2\right)<b^2\\ .
\label{cond2}
\eeq
Finally, we require the physical masses to be positive:
\beq
m_{h^0_1}^2>0\ ,\qquad m_{A_1}^2>0\, ,
\label{cond3}
\eeq
where $m_{h^0_1},\,m_{A_0}$ are defined in Eqs. (\ref{mhs}),(\ref{mBs}) and (\ref{mg}). Without loss of generality one can choose $0<\beta<\pi/2$. After plugging Eqs. (\ref{v1}), (\ref{v2}) and (\ref{v3}) in (\ref{cond1}), (\ref{cond2}) and (\ref{cond3}) all these conditions are satisfied for 
\beq
0<b<\frac{t_{2\beta}}{16}\left(g^2_L+g^2_Y\right) \left(v_{N4}^2-c_{2\beta} v_H^2\right),\quad c_{2 \beta} v_H^2 < v^2_{N4},\quad 0<\beta<\frac{\pi}{4}\ ,
\eeq
or
 \beq
 b>0,\qquad \pi/4\leqslant\beta<\pi/2\ .
 \label{vcond}
 \eeq 
We will investigate the parameter space defined by the conditions (\ref{vcond}) since the top mass is more easily accomodated in this region.

\section{Mass Spectrum}
\label{MSp}
The model conserves lepton and baryon numbers $L$ and $B$. After the EWSB, the Lagrangian is invariant under the residual $U(1)_{EM}\times U(1)_{N4}$. We can therefore write the gauge boson, fermion, and scalar (squared) mass matrices in block diagonal form in the basis of EM- and N4-charges and L and B numbers.

The mass matrices of all the SM fermions and their superpartners are of the same form as those obtained in the MSSM; these can be found for example in \cite{Martin:1997ns}. The EW gauginos, Higgs scalar doublets and their superpartners mix with the \nf sector.  
Finally the fields $N_L, \bar{N}_R$, and their scalar superpartners will not mix at tree level with other SM fields with EM charge $Q_{EM}=1$ (where we defined $Q_{EM}=T^3_L+Y$). 

\subsection{Gauge Bosons}\label{GBm}
After EWSB some 4SYM gluons and EW gauge bosons acquire mass. The gauge sector in the model Lagrangian can be written as a function of the mass eigenstates as: 
\beq
-{\cal L}_{g\textrm{-}mass}=g_{N4}^2 v_{N4}^2 G^{+}_\mu G^{-\mu}+\frac{g^2_L}{2} \left( v_{N4}^2+ v_H^2\right) W_\mu^+ W^{-\mu}+\frac{g^2_L+g^2_Y}{4} \left( v_{N4}^2+ v_H^2\right) Z_\mu Z^\mu
\label{gmL}
\eeq
where
\beq
G^{\pm}_\mu=\frac{1}{\sqrt{2}}\left( G^1_\mu \mp i\, G^2_\mu \right)\ ,\ W^{\pm}_\mu=\frac{1}{\sqrt{2}}\left( W^1_\mu \mp i\, W^2_\mu \right)\ ,\  Z_\mu=c_w W^3_\mu-s_w B\ ,\ t_w=\frac{g_Y}{g_L}.
\label{gme}
\eeq
The $\pm$ exponent of the 4SYM gluon refers to the $U(1)_{N4}$ charge, while the $\pm$ exponent on the EW gauge bosons refer to the usual EM charge. The remaining, massless states are the 4SYM photon and the EW photon:
\beq
G_\mu=G^3_\mu\ ,\qquad  A_\mu=s_w W^3_\mu + c_w B\ .
\label{g0e}
\eeq
The phenomenological constraints on a new  $U(1)$ massless gauge boson were studied in \cite{Dobrescu:2004wz}, and their analysis shows that the operators coupling such photon to the SM fields needs to be suppressed by scales at least of the order of the EW scale. This could provide relevant constraints on our model.
 
The tree-level masses of $G,\ W$ and $Z$ can be read off from Eq.(\ref{gmL}):
\beq
\label{mg}
m_G=g_{N4} v_{N4}\ ,\qquad  m_W= \frac{g_L}{2}\sqrt{v_{N4}^2+v_H^2}\ ,\qquad m_Z= \frac{m_W}{c_w}\ .
\eeq
From these masses and the eigenstates in Eq.(\ref{gme}) it is immediate to evaluate the EW oblique parameters at tree level by using the formulas in \cite{Burgess:1993vc}: we find $S=T=U=0$ at tree level. 

\subsection{Fermions}\label{Fm}
The charge of each fermion under the unbroken group $U(1)_{EM}\times U(1)_{N4}$ is indicated by the superscript of the form $(x,y)$, where $x$ denotes the charge under $U(1)_{EM}$ and $y$ denotes charge under $U(1)_{N4}$. The Weyl fermion mass terms are:
\be
-{\cal L}_{f\textrm{-}mass}
&=&
\frac{1}{2}\left(
\chi^{(0,0)}\right)^T {\cal M}_{(0,0)}\chi^{(0,0)}
+\left(\chi^{(0,+)}\right)^T {\cal M}_{(0,+)}\chi^{(0,-)}
+\left(\chi^{(+,0)}\right)^T {\cal M}_{(+,0)}\chi^{(-,0)}
\nonumber\\
&+&
m_{(+,+)}\chi^{(+,+)}\chi^{(-,-)} 
+m_{(+,+)}^*\chi^{(+,-)}\chi^{(-,+)}
+m_{(++,0)} \chi^{(++,0)}\chi^{(--,0)} +c.c.\ ,
\ee
where
\begin{align}
\chi^{(0,0)}&=\left(H_2,H^\prime_1,\tilde{W}_3,\tilde{B},D_L^3,\bar{D}_R^3\right)\,,& 
\chi^{(0,\pm)}&=\left(\frac{D_L^1\mp i\, D_L^2}{\sqrt{2}},\frac{\bar{D}_R^1\mp i\, \bar{D}_R^2}{\sqrt{2}}\right),\nonumber\\ 
\chi^{(+,0)}&=\left(H_1,\frac{\tilde{W}_1- i\, \tilde{W}_2}{\sqrt{2}},U_L^3,\bar{N}_R\right)\,,&\chi^{(-,0)}&=\left(H^\prime_2,\frac{\tilde{W}_1+ i\, \tilde{W}_2}{\sqrt{2}},\bar{U}_R^3,N_L\right),\nonumber\\ 
\chi^{(+,\pm)}&=\frac{U_L^1\mp i\, U_L^2}{\sqrt{2}},\quad \chi^{(-,\pm)}=\frac{\bar{U}_R^1\mp i\, \bar{U}_R^2}{\sqrt{2}} \ ,  &\chi^{(++,0)}&=\bar{E}_R,\ \chi^{(--,0)}=E_L,
\end{align}
and, at tree-level,
\be
\label{mn}
{\cal M}_{(0,0)}=\frac{1}{2}\left(\begin{array}{cccccc} 0 & -2\mu & i s_\beta g_L v_H & -i s_\beta g_Y v_H & 0 & 0 \\ 
-2\mu & 0 & -i c_\beta g_L v_H & i c_\beta g_Y v_H & 0  & 0\\
i s_\beta g_L v_H & -i c_\beta g_L v_H & 2 M_{\tilde{W}} & 0 & i g_L v_{N4}  & 0\\
-i s_\beta g_Y v_H & i c_\beta g_Y v_H & 0 & 2 M_{\tilde{B}} & -i g_Y v_{N4}  & 0\\
0 & 0 & i g_L v_{N4} & -i g_Y v_{N4} & 0  & 0\\
 0  & 0  & 0  & 0  & 0  & 2 M_{D} \end{array}\right)\ ,
\ee
\be
\label{mtc}
{\cal M}_{(0,+)}=\left(\begin{array}{cc}0 & i g_{N4} v_{N4} \\-i g_{N4} v_{N4} & M_D \end{array}\right)\ ,
\ee
\be
\label{mc}
{\cal M}_{(+,0)}=\frac{1}{\sqrt{2}}\left(\begin{array}{cccc}\sqrt{2}\mu & -i s_\beta g_L v_H & -y_U v_{N4} & 0 \\
-i c_\beta g_L v_H & \sqrt{2} M_{\tilde{W}} & 0 & 0\\
0 & -i g_L v_{N4} & y_U s_\beta v_H  & 0 \\0 & 0 & 0 & y_N s_\beta v_H\end{array}\right)\ ,
\ee
\beq
\label{mcc}
m_{(+,+)}=-i g_{N4} v_{N4}+ \frac{y_U s_\beta v_H}{\sqrt{2}}\ , \qquad  m_{(++,0)}= \frac{y_E c_\beta v_H}{\sqrt{2}} .
\eeq
The star indicates complex conjugation, while a tilde indicates the fermion superpartner of the corresponding gauge boson. $M_{\tilde{W}}$ and $M_{\tilde{B}}$ correspond to the wino and the bino soft masses, respectively. It is important for the phenomenological bounds to notice that, from the last equation, and the mass of the top: $m_t=y_t s_\beta v_H/\sqrt{2}$, it follows that 
\beq
m_t=\frac{y_t}{y_E}t_\beta m_{(++,0)}\ .
\label{mt}
\eeq
 
The squared masses of each type of fermions are obtained by diagonalizing the corresponding ${\cal M} {\cal M}^\dagger$. We note that $\bar{D}^3_R$ has become the gaugino of the residual $U(1)_{N4}$ with mass $M_D$. For illustration we provide the explicit form of the $(0,+)$ fermion masses obtained diagonalizing the seesaw-like matrix in Eq.(\ref{mtc}):
\beq
\label{mtc01}
m_{(0,+)}=\sqrt{\frac{M^2_D}{4}+g_{N4}^2 v_{N4}^2}\pm \frac{M_D}{2}.
\eeq
This also shows that in order for all matter fields to become massive the vev $v_{N4}$ must be nonzero.

\subsection{Scalars}
\subsubsection{Tree-Level}\label{SmTL}
The complete potential is given by
\be
V=V_{N4}+V_{MSSM},\qquad V_{N4}=-{\cal L}_{D}-{\cal L}_{F}-{\cal L}_{soft}
-\left(\frac{1}{2}M_D\bar{D}^a_R\bar{D}^a_R+c.c.\right)
\ ,
\ee
where $V_{MSSM}$ can be found in \cite{Martin:1997ns}, while ${\cal L}_{D}, {\cal L}_{F}$, and ${\cal L}_{soft}$, are given in Appendix \ref{appxA}. The SM squarks and sleptons do not mix at tree-level with the \nf scalars or heavy new scalar leptons and therefore their mass spectrum assumes the same form as in the MSSM. The Higgs scalar fields, $\tilde{H}$ and  $\tilde{H}^\prime$, on the other hand, mix with the \nf scalars. The squared mass matrices of the CP-even and -odd EM neutral Higgs scalars are given by, respectively,
\be
\label{mh2}
{\cal M}^2_h&=&\frac{1}{4}
\left(
\begin{array}{ccc}
 \left(g_L^2+g_Y^2\right) s_{\beta }^2 v_H^2+4 b t^{-1}_{\beta } & -c_{\beta } \left(g_L^2+g_Y^2\right) s_{\beta } v_H^2-4 b & \left(g_L^2+g_Y^2\right) s_{\beta } v_H v_{N4}
   \\
 -c_{\beta } \left(g_L^2+g_Y^2\right) s_{\beta } v_H^2-4 b & c_{\beta }^2 \left(g_L^2+g_Y^2\right) v_H^2+4 b t_{\beta } & -c_{\beta } \left(g_L^2+g_Y^2\right) v_H v_{N4} \\
 \left(g_L^2+g_Y^2\right) s_{\beta } v_H v_{N4} & -c_{\beta } \left(g_L^2+g_Y^2\right) v_H v_{N4} & \left(g_L^2+g_Y^2\right) v_{N4}^2
\end{array}
\right)\,,
\nonumber\\
\left({\cal M}^2_h\right)_{ij}&=&\left.\frac{\partial^2 V}{\partial \phi^h_i \partial \phi^h_j}\right|_{\phi=\left<  \phi \right>}\, ,\ \phi^h=\Re\left( \tilde{H}_2, \tilde{H}_1^\prime, \tilde{D}_L^3 \right)\, ,
\ee
and
\be
\label{mA2}
{\cal M}^2_A&=&\left(
\begin{array}{ccc}
 b t^{-1}_{\beta } & b & 0 \\
 b & b t_{\beta } & 0 \\
 0 & 0 & 0
\end{array}
\right)\, ,\  \left({\cal M}^2_A\right)_{ij}=\left.\frac{\partial^2 V}{\partial \phi^A_i \partial \phi^A_j}\right|_{\phi=\left<  \phi \right>}\, ,\quad \phi^A=\Im\left( \tilde{H}_2, \tilde{H}_1^\prime, \tilde{D}_L^3 \right)
\ee
From Eqs. (\ref{mh2}) and (\ref{mA2}) the squared masses of the CP-even and -odd Higgs scalars are 
\beq\label{mhs}
m^2_{h^0_0}=m_{A_0}^2=0,\ m^2_{h^0_{1,2}}=\frac{1}{2} \left(m_{A_1}^2+m_Z^2\mp\sqrt{\left(m_{A_1}^2-m_Z^2\right)^2+4 m_{A_1}^2 m_B^2 s_{2 \beta }^2}\right),\ m^2_{A_1}=\frac{2 b}{s_{2\beta}},
\eeq
where we have defined the quantity
\beq\label{mBs}
m_B^2=\frac{g_Y^2+g_L^2}{4} v_H^2
\eeq
which does not correspond to the mass of any particle. In the limit $v_{N4}=0$, however, $ m_B=m_Z$ and one recovers the MSSM results for the masses of the CP-even Higgs scalars.  

The massless eigenstates $h^0_0,\ \pi_Z$ (the longitudinal degree of freedom of the $Z$ boson), and $A_0$, are expressed by
\begin{align}
h^0_0=&N_h\left( s_\beta v_{N4},  c_\beta v_{N4},  c_{2\beta} v_H \right)\cdot\phi^h\,,& N_h^{-2}&=v^2_{N4} +  c^2_{2\beta} v^2_H\,,\\
\pi_Z=&N_Z \left( s_\beta v_{H},  -c_\beta v_{H},  v_{N4} \right)\cdot\phi^A\,,& N_Z^{-2}&=v^2_{N4} + v^2_H\,,\label{espiZ}\\
A_0=&N_A \left( s_\beta v_{N4},  -c_\beta v_{N4},  -v_{H} \right)\cdot\phi^A\,,& N_A^{-2}&=v^2_{N4} + v^2_H\ ,
\end{align}
with $\phi^{h,A}$ defined respectively in Eqs.~(\ref{mh2},\ref{mA2}).
The masslessness of $h^0_0$ and $A_0$  will not survive at the one-loop level.

The remaining scalar squared mass matrices are given in Appendix \ref{appMs2}. By using these results and those given in Eqs. (\ref{mg}), (\ref{mn}), (\ref{mh2}) and (\ref{mA2}),
we have checked that the SUSY invariant contributions to the supertrace of the squared mass matrices cancel out, as they should. 

\subsubsection{One-Loop}
We calculate the one-loop contributions to the masses of the CP-even and -odd neutral (both under $U(1)_{EM}$ and $U(1)_{N4}$) scalars. We expect the lightest eigenstates, $h^0_0$ and $A_0$, that are accidentally massless at tree level, to receive non-zero contributions to their masses from the one-loop effective potential. The one loop effective potential is  \cite{Coleman:1973jx}:
\begin{align}
\label{CWV}
\Delta V_1=\frac{1}{64\pi^2}
S\tr\left[ \mathcal{M}^4\left( \phi \right)
\left(\ln\frac{\mathcal{M}^2\left( \phi \right)}{\mu_r^2}-\frac{3}{2}\right)+2\mathcal{M}^2\left( \phi \right) \mu_r^2 \right],
\end{align}
where $\mathcal{M}^2\left( \phi \right)$ are field-dependent mass matrices not evaluated at their vevs, defined by:
\beq
\left({\cal M}^2\left( \phi \right) \right)_{ij}=\frac{\partial^2 V}{\partial \phi_i \partial \phi_j}\,,
\eeq
 and $\mu_r$ is the renormalization scale. The last term in Eq.(\ref{CWV}) renormalizes the one-loop contributions to the scalar masses to zero when $\mu_r^2=\mathcal{M}^2\left( \left<\phi\right> \right)$.\footnote{In case there is more than one field, one should use different scales $\mu_r$ for each contribution to the supertrace to get an exactly vanishing one-loop correction to the mass.} This term gives a very small contribution to $\Delta V_1$ since it arises only from the SUSY breaking terms which are generally small to avoid a large fine tuning. Therefore we neglect it. To minimize the correction from higher order contributions to V, we take $\mu_r$ equal to the mass of the heaviest particle among the eigenstates presented in Sections \ref{GBm}, \ref{Fm}, and \ref{SmTL}

The one-loop mass matrix correction, $\Delta\mathcal{M}^2_{a}$, for any real field $a$ with $n$ components can be
extracted from $\Delta V_1$ by numerically evaluating the derivatives of the mass eigenvalues
with respect to the fields evaluated on the vevs \cite{Elliott:1993bs}, where
\begin{align}
(\Delta\mathcal{M}^2_{a})_{ij}
&=\left.\frac{\partial^2 \Delta V_1(a)}{\partial a_i\partial a_j}\right|_{a=\left<  a \right>}
+\Delta M^2_{ij}\,,
\label{eq:CWex02}\\
\left.\frac{\partial^2 \Delta V_1(a)}{\partial a_i\partial a_j}\right|_{a=\left<  a \right>}&=\sum\limits_{k}\frac{1}{32\pi^2}
\frac{\partial m^2_k}{\partial a_i}
\frac{\partial m^2_k}{\partial a_j}
\left.\ln\frac{m_k^2}{\mu_r^2}\right|_{a=\left<  a \right>}
+\sum\limits_{k}\frac{1}{32\pi^2}
m^2_k\frac{\partial^2 m^2_k}{\partial a_i\partial a_j}
\left.\left(\ln\frac{m_k^2}{\mu_r^2}-1\right)\right|_{a=\left<  a \right>}\,,\nonumber\\
\Delta M^2_{ij}&=-\frac{\delta_{ij}}{\phi^h_i}\left.\frac{\partial \Delta V_1(\phi^h)}{\partial \phi^h_i}\right|_{\phi^h=\left< \phi^h \right>}\nonumber\\
&=-\sum\limits_{k}\frac{1}{32\pi^2}m^2_k
\frac{\delta_{ij}}{\phi^h_i}
\frac{\partial m^2_k}{\partial \phi^h_i}
\left.\left(\ln\frac{m_k^2}{\mu_r^2}-1\right)\right|_{\phi^h=\left< \phi^h \right>}\,.
\label{eq:CWex03}
\end{align}
The second term in Eq.(\ref{eq:CWex02}) takes into account the shift in
the minimization conditions (see \cite{Elliott:1993bs}), and $m^2_k$ is the set of
mass eigenvalues of the field dependent mass matrix  $\mathcal{M}^2\left( \phi \right)$. Notice that $\Delta M^2_{ij}$ has to be included in the expression of $(\Delta\mathcal{M}^2_{a})_{ij}$ only when $a_i$ are the CP-even or -odd Higgses, since $\Delta M^2_{ij}$ gives the shift of the soft mass parameters of the scalar fields that develop a non-zero vev. The Goldstone bosons do not contribute to $\Delta\mathcal{M}^2_{a}$.

{ In this first estimate we compute $\Delta\mathcal{M}^2$ for the neutral higgses neglecting the contributions from top and stop loops. We consider the fields given in Table \ref{MSCTsuperfields}, plus the $W$ and $B$ bosons and their superpartners. In this way the supertrace  receives contributions only from the soft mass terms. We therefore consider our results for the one-loop masses of the CP-even and -odd Higgses an estimate of the values that can be obtained when taking into account the full spectrum of the model.}

It is seen that except for the ordinary EM neutral Goldstone boson, which can be interpreted as the longitudinal component of the $Z$ boson, no other neutral scalar is massless. 
The mass of the lightest physical states, $h^0_0$ and $A_0$, has a strong dependence on the size of the Yukawa couplings in the superpotential, Eq.(\ref{spmwt}). A random scan of the parameter space, with the constraint that the SUSY breaking scale, given in Eqs.(\ref{Lsof},\ref{vin}), is around the TeV region and with $\pi/4<\beta<\pi/2$, gives, before adding the one loop corrections from the MSSM sector, these rough estimates:
\begin{align}
\label{1Lmh0}
m_{h_0^0}\sim 10 \textrm{ GeV}\,,  \qquad  g_{N4}=y_U=y_N=y_E=y_R=1\,, \nonumber\\
m_{h_0^0} \sim 125  \textrm{ GeV}\,, \qquad   g_{N4}=y_U=y_N=y_E=y_R=\pi\,.
\end{align}
We have also tried to reach a larger value of the masses by optimizing the search around the maximum value of the initial sample of parameters and obtain in this case $m_{h_0^0}^{max}\sim 30$~GeV and $m_{h_0^0}^{max}\sim 270$~GeV for the same choice of Yukawas above. Another major contribution arises from including the top and stop loops. We estimate it to give around $30$ GeV additional contribution.

The mass of $A_0$ for the parameter values that maximize $m_{h_0^0}$ is $m_{A_0}=8\,(27)$ GeV for $g_{N4}=...=1\,(\pi)$. Nota bene that this does not imply the mass of $A_0$ has to be this light, since we have not maximized its mass through a parameter scan. Moreover, we find that $m_{A_0}$ is proportional to $a_{N4}$, which we constrained to be smaller than two TeV. Consequentially, the mass of $A_0$ can be easily increased by increasing $a_{N4}$.

In the next section we impose the experimental bounds on the mass spectrum to discuss its phenomenological viability, and use the renormalization group equations to determine the perturbative range of our results.

\section{Phenomenological Viability}
\label{PhV}
The lower bounds on the mass of the lightest neutralino and chargino are \cite{Amsler:2008zzb}:
\beq
\label{exmass}
m_{\chi_0^0}>46 \textrm{ GeV}\,,\ m_{\chi_0^\pm}>94 \textrm{ GeV}\,. 
\eeq
These limits refer to the MSSM, but are rather general, since they are extracted mostly from the $Z$ decay to neutralino-antineutralino pair the former, and from photo-production of a chargino-antichargino pair at LEPII the latter. We can therefore assume these limits to hold also for our model. Because of their generality and independence from the coupling strength (as long as it is not negligible), we use the lower bound on the chargino mass also for the mass of the doubly-charged chargino $E$. The presence of the term proportional to $y_R$ in the superpotential, Eq.(\ref{spmwt}) allows it to decay into singly charged ordinary particles. Therefore it escapes cosmological constraints on charged stable particles. The electrically neutral 4SYM gaugino, $\bar{D}^3_R$ with mass $M_D$, is an EW singlet fermion, analogous to a right-handed neutrino, and hence can be very light. Because of this, and to keep the lightest ${(0,\pm)}$ fermion, Eq.(\ref{mtc01}), massive enough, we assume $M_D\ll g_{N4} v_{N4}$.

\begin{figure}
\begin{center}
\includegraphics[width=3.5in]{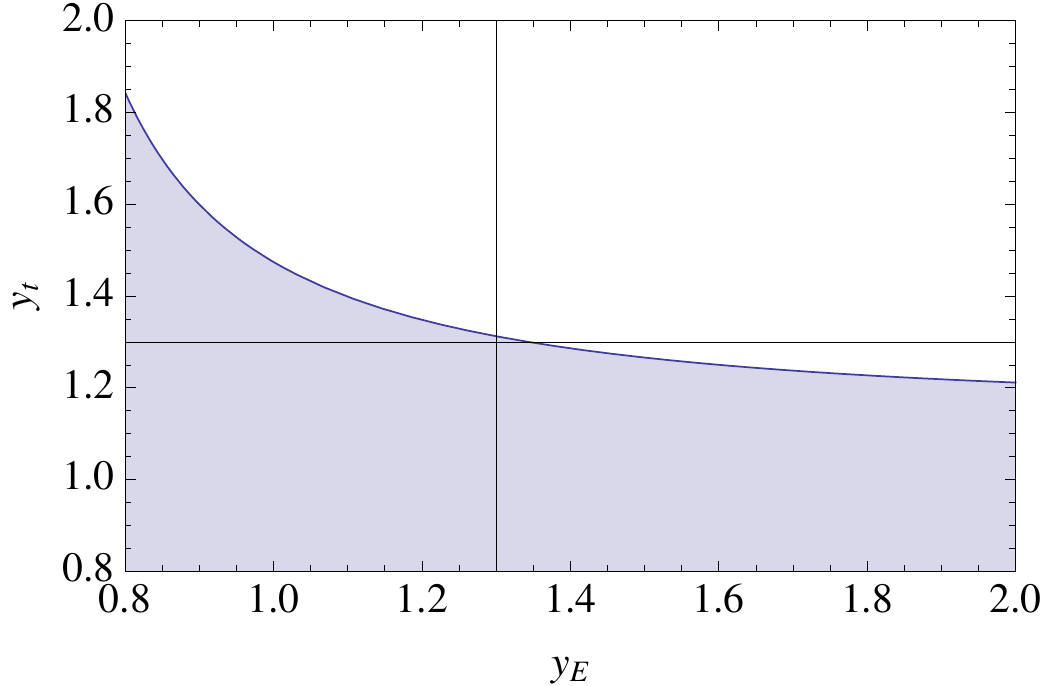}
\end{center}
\caption{Shaded area shows experimentally excluded values of the Yukawa couplings $y_t$ and $y_E$.}
\label{fig:ytyE}
\end{figure}

Other useful limits on the parameters are obtained by using the fact that the smallest eigenvalue of a semi-positive definite square matrix is smaller or equal to any eigenvalue of the principal submatrices. From the absolute square of the ${(0,0)}$ fermion mass matrix, Eq.(\ref{mn}), we get
\be
\label{netb}
M_{\tilde{B}}^2&>&\left(46 \textrm{ GeV}\right)^2-\frac{g_Y^2}{4} \left(v_H^2+v_{N4}^2\right)=\left(13.5 \textrm{ GeV}\right)^2\,,\ \mu>46 \textrm{ GeV}\,,\nonumber\\
v_{\text{N4}}&>&2 \frac{46\textrm{ GeV}}{\sqrt{g_L^2+g_Y^2}}=124\textrm{ GeV}\,,\ v_H<213\textrm{ GeV}\,,
\ee
where we used, from Eq.(\ref{mg}),
\beq
\sqrt{v^2_{H}+v_{N4}^2}=246 \textrm{ GeV}.
\eeq
From the  ${(+,0)}$ and ${(++,0)}$ fermion mass matrices, Eqs.(\ref{mc},\ref{mcc}), we get
\beq
\label{chab}
M_{\tilde{W}}^2>\left(94 \textrm{ GeV}\right)^2-\frac{1}{2} c_{\beta }^2 g_L^2 v_H^2=\left(63.5 \textrm{ GeV}\right)^2\,,\ \frac{y_E c_{\beta } v_H}{\sqrt{2}}>94 \textrm{ GeV}\,.
\eeq
From Eq.(\ref{mt}), with $m_t=173$ GeV, and the bounds (\ref{netb},\ref{chab}), it follows that
\beq
\label{ytyE}
y_t > \frac{173}{213} \sqrt{\frac{1}{\frac{1}{2}-\frac{94^2}{y_E^2 213^2}}}\,.
\eeq
This last bound is plotted in Figure \ref{fig:ytyE}, where the shaded area shows the values of $y_t$ and $y_E$ excluded by the experiment: it is evident from the plot in Figure \ref{fig:ytyE} that either $y_t$ or $y_E$ is constrained to be larger than about 1.3.\footnote{Had we chosen the hypercharge parameter y=-1 rather than 1, the constraints in Eqs.(\ref{netb},\ref{chab},\ref{ytyE}) would be the same with $y_E$ and $y_N$ interchanged. although a more detailed study would be necessary, we expect that the choice $y=-1$ produces the same general results and conclusions that we present in this paper for y=1.}

To further study the phenomenological viability of the spectrum we now analyze the evolution of the couplings using the two-loop renormalization group equations (RGE) given in Appendix \ref{rge}. In this calculation we assume a generic Yukawa coupling $y_{N4}$ in place of $g_{N4}$ in Eq.(\ref{spmwt}).

We find that typical behavior for the phenomenologically favored large Yukawa couplings is that the up-type couplings flow to an ultraviolet fixed point, while the remaining ones, such as $y_{N4}$, flow toward zero. Qualitatively, the UV fixed point is caused by the large anomalous dimension of the up-type Higgs, which results from its coupling to the 4SYM sector. The fixed point behavior begins rather quickly, as the largest couplings reach their fixed point value $y_\star\simeq 6$ at around 2 TeV. At two-loops the 4SYM gauge coupling $g_{N4}$ decreases as a function of increasing scale, but the evolution is very slow in comparison to the Yukawa couplings. We also find that at two loops $g_{N4}=y_{N4}$ is an infrared fixed point, in agreement with the findings in \cite{Petrini:1997kk}.

\begin{figure}[h]
\begin{center}
\includegraphics[width=4.5in]{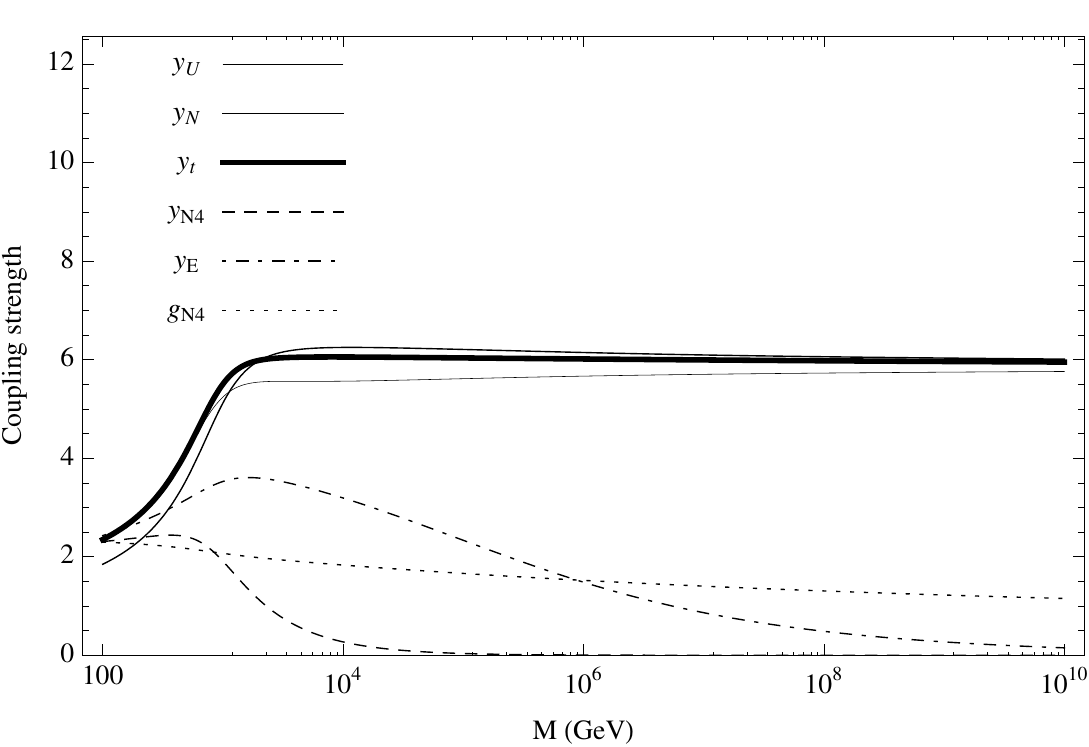}
\end{center}
\caption{Plot of $g_{N4},y_{N4},y_U,y_t,y_N,y_E$ as a function of the renormalization scale $M$: the couplings are normalized for $M=m_Z$ to $y_N=1.8,\,g_{N4}=y_{N4}=y_U=y_t=2.3,\,y_E=2.4$.}
\label{fig:RGE}
\end{figure}

We scanned the parameter space of
the model for Yukawa couplings that delay the onset of the fixed point, while satisfying the neutralino and chargino mass limits, Eq. (\ref{exmass}), and maximizing the mass of the CP even Higgs scalar. 
The dimensionful soft SUSY breaking parameters were taken to be around a TeV. In Figure~\ref{fig:RGE} are plotted $g_{N4},y_{N4},y_U,y_t,y_N,y_E$ as a function of the renormalization scale $M$: the couplings are normalized for $M=m_Z$ to $y_N=1.8,\,g_{N4}=y_{N4}=y_U=y_t=2.3,\,y_E=2.4$. Summarizing, $g_{N4}$ runs towards zero in the ultraviolet, while the Yukawa couplings $y_U,\,y_N,\,y_t$, responsible for the mass of the heavy upper components of weak doublets, increase and flow close to an ultraviolet fixed point at around 2 TeV. 
For such values of the Yukawa couplings we can achieve the following spectrum: 
\beq
\label{maxLEP}
m_{\chi^0_0}=47 \textrm{ GeV}\,,\ m_{\chi^\pm_0}= 96\textrm{ GeV}\,,\ m_{h_0^0}=95\textrm{ GeV}\,,\  m_{A_0}=32\textrm{ GeV}\,. 
\eeq
The spectrum above represents a sample point. For example the value of the $A_0$ mass can be higher. It can also be increased by including  the stop and top loops when determining the one loop effective potential. Another way to increase the mass of $A_0$ is by simply increasing the value of $a_{N4}$.  

Assuming the spectrum in \eqref{maxLEP}  at a $e^+e^-$ collider, the main production channel of the $A_0$ would be via  $Z\rightarrow h_0^0A_0$. For a hadron collider, one has also production via gluon-gluon fusion and associated production with heavy quarks. To determine these processes one needs the following  couplings:
\be
g_{h_0^0 A_0 Z}&:&
-\frac{\sqrt{g_Y^2+g_L^2}}{2}\frac{c_{2\beta}\sqrt{v_{N4}^2+v_H^2}}{\sqrt{v_{N4}^2+v_H^2c_{2\beta}^2}}\\
g_{A_0 \bar{b}\gamma_5 b}&:&
-\frac{m_b}{\sqrt{v_{N4}^2+v_H^2}}\frac{v_{N4}}{v_H}\\
g_{A_0 \bar{t}\gamma_5 t}&:&
\frac{m_t}{\sqrt{v_{N4}^2+v_H^2}}\frac{v_{N4}}{v_H}
\ee
where $m_{f}$ is the fermion mass. The formulae are generic for up and down type fermions.

For $\beta\sim\pi/4$ we find $g_{h_0^0 A_0 Z}\sim 0$ implying that, compared to the MSSM, there is a depletion of the $A_0$ production rate at  $e^+e^-$ colliders. As for the constraints from hadron colliders, with $\tan\beta \simeq v_{N4}/v_H \simeq1$, our model's couplings to quarks are of the same order of the MSSM couplings and therefore the model parameter space has not yet been entirely constrained by the LHC.  This simple analysis shows that the model is not yet ruled out.  

\section{Conclusions and Outlook}

We have investigated a perturbative \nf sector coupled to the MSSM. The SUSY breaking scale is constrained by naturalness requirements to be around the TeV scale. First we showed that the model allows for a stable vacuum, in which the EW symmetry is broken by expectation values of the MSSM Higgses and scalars of 4SYM. We then reduced the parameter space by imposing naturalness of the couplings and masses, one loop vacuum stability, perturbativity at the EW scale, and experimental constraints.

Because of the additional vev of the 4SYM scalar sector, which contributes to the masses of the EW gauge bosons, but not to that of quarks and leptons, all Yukawa couplings are larger than in the MSSM. By running the two loop renormalization group equations of the dimensionless couplings, we found that the Yukawa of the heavy up-type fermions flow to a common UV fixed point at about 2 TeV. The remaining couplings decrease with increasing energy.

There are many possible interesting signatures of this model for collider experiments. Compared to the MSSM, our model features several new states, such as doubly charged particles, and several light scalars. In the future we plan to explore the processes relevant for collider experiments, as well as dark matter phenomenology, which will be substantially different than in the MSSM.

Since our model features a new ${\cal N}=4$ sector at the EW scale, collider experiments have the possibility to explore string theory directly. This is because the new scalars coming from this sector can be directly identified with the {\it extra} six space coordinates of ten dimensional supergravity.  This link is even more clear when considering the \nf sector in the nonperturbative regime which can be investigated using AdS/CFT techniques.

\acknowledgments
We would like to thank Matti J\"arvinen for useful discussions, and R. Sekhar Chivukula for valuable comments. 

\newpage
\appendix

\section{MSCT Lagrangian}
\label{appxA}
The Lagrangian of a supersymmetric theory can, in general, be defined by
\beq
{\cal L}={\cal L}_{kin}+{\cal L}_{g-Yuk}+{\cal L}_{D}+{\cal L}_{F}+{\cal L}_{P-Yuk}+{\cal L}_{soft},
\label{LP}
\eeq
where the labels refer to the kinetic terms, the Yukawa ones given by gauge and superpotential interactions, the $D$ and $F$ scalar interaction terms, and the soft SUSY breaking ones. All these terms can be expressed in function of the elementary fields of the theory with the help of the following equations:
\be
{\cal L}_{kin}&=&-\frac{1}{4} F_j^{\mu\nu a}F_{j\mu\nu}^a-i \bar{\lambda}_j^a\bar{\sigma}^\mu D_\mu \lambda_j^a-D^\mu \phi^{a\dagger}_i D_\mu \phi^a_i - i \bar{\chi}^a_i\bar{\sigma}^\mu D_\mu\chi^a_i\ ,\label{comp1}\\
{\cal L}_{g-Yuk}&=&\sum_j i\sqrt{2} g_j \left( \phi_i^{\dagger} T^a_j\chi_i\lambda^a_j-\bar{\lambda}^a_j\bar{\chi}_i T^a_j \phi_i \right)\ ,\label{comp2}\\
{\cal L}_{D}&=&-\frac{1}{2}  \sum_j g_j^2 \left(\phi_i^{\dagger} T^a_j \phi_i \right)^2\ ,\label{comp3}\\
{\cal L}_{F}&=&-\left|\frac{\partial P}{\partial \phi^a_i} \right|^2\ ,\label{comp4}\\
{\cal L}_{P-Yuk}&=&-\frac{1}{2}\left[\frac{\partial^2 P}{\partial \phi^a_i\partial \phi^b_l}\chi^a_i\chi^b_l+h.c. \right]\label{comp5},
\ee
where $i,l$ run over all the scalar field labels, while $j$ runs over all the gauge group labels, and $a,b$ are the corresponding gauge group indices. Furthermore, we normalize the generators in the usual way, by taking the index $T(F)=\frac{1}{2}$, where 
\[
\tr T^a_RT^b_R=T(R)\delta^{ab},
\]
with $R$ here referring to the representation ($F$=fundamental).
The SUSY breaking soft terms, moreover, are obtained by re-writing the superpotential in function of the scalar fields alone, and by adding to it its Hermitian conjugate and the mass terms for the gauginos and the scalar fields.

We refer to \cite{Martin:1997ns} and references therein for the explicit form of ${\cal L}_{MSSM}$ in terms of the elementary fields of the MSSM, and focus here only on ${\cal L}_{N4}$. The kinetic terms are trivial and therefore we do not write them here. The gauge Yukawa terms are given by
\be
{\cal L}_{g-Yuk}&=& \sqrt{2}g_{N4}\left(  \tilde{\bar{U}}_L^bU_L^c\bar{D}^a_R-D_R^a\bar{U}_L^b\tilde{U}_L^c + \tilde{\bar{D}}_L^bD_L^c\bar{D}^a_R-D_R^a\bar{D}_L^b\tilde{D}_L^c
+ \tilde{U}_R^b\bar{U}_R^c\bar{D}^a_R-D_R^aU_R^b\tilde{\bar{U}}_R^c  \right)\epsilon^{abc}\nonumber\\
&+&i\frac{g_L}{\sqrt{2}} \left(  \tilde{\bar{Q}}_L^i Q_L^j\tilde{W}^k-\tilde{\bar{W}}^k \bar{Q}_L^i \tilde{Q}_L^j+\tilde{\bar{L}}_L^i L_L^j\tilde{W}^k-\tilde{\bar{W}}^k \bar{L}_L^i \tilde{L}_L^j  \right)\sigma^k_{ij}\nonumber\\
&+&i\sqrt{2} g_Y\sum_p Y_p\left(  \tilde{\bar{\chi}}_p\chi_p\tilde{B}-\tilde{\bar{B}}\bar{\chi}_p\tilde{\chi}_p  \right),\ \chi_p=U^a_L,D^a_L,\bar{U}^a_R,N_L,E_L,\bar{N}_R,\bar{E}_R\ ,
\ee
where $\tilde{W}^k$ and $\tilde{B}$ are respectively the wino and the bino, $\sigma^k$ the Pauli matrices, $i,j=1,2;\ k,a,b,c=1,2,3$; and the hypercharge $Y_p$ is given for each field $\chi_p$ in Table \ref{MSCTsuperfields}.

The $D$ terms are given by
\beq
{\cal L}_D=-\frac{1}{2}\left( g_{N4}^2 D^a_{N4} D^a_{N4}+ g_{L}^2 D^k_{L} D^k_{L}+ g_{Y}^2 D_{Y} D_{Y}\right)+\frac{1}{2}\left( g_{L}^2 D^k_{L} D^k_{L}+ g_{Y}^2 D_{Y} D_{Y}\right)_{MSSM},
\eeq
where
\be
D^a_{N4}&=&-i \epsilon^{abc}\left(  \tilde{\bar{U}}_L^b\tilde{U}^c_L+ \tilde{\bar{D}}_L^b\tilde{D}^c_L+ \tilde{U}_R^b\tilde{\bar{U}}^c_R  \right),\ D^k_{L}=\frac{\sigma^k_{ij}}{2} \left(  \tilde{\bar{Q}}_L^{i\,a}\tilde{Q}^{j\,a}_L + \tilde{\bar{L}}_L^i\tilde{L}^j_L  \right) +D^k_{L,MSSM}\nonumber\\
D_{Y}&=&\sum_p Y_p \tilde{\bar{\chi}}_p\tilde{\chi}_p+D_{Y,MSSM}.
\ee
In these equations the $D^k_{L, MSSM}$ and $D_{Y, MSSM}$ auxiliary fields are assumed to be expressed in function of the MSSM elementary fields \cite{Martin:1997ns}.
The rest of the scalar interaction terms\footnote{We consider the constants in the superpotential to be real to avoid the contribution of CP violating terms.} is given by
\be
{\cal L}_F&=&-g_{N4}^2\left[ \left(  \tilde{U}_L^b\tilde{\bar{U}}_L^b+\tilde{D}_L^b\tilde{\bar{D}}_L^b+\tilde{\bar{U}}_R^b\tilde{U}_R^b\right)^2- \left(  \tilde{U}_L^b\tilde{\bar{U}}_L^c+\tilde{D}_L^b\tilde{\bar{D}}_L^c+\tilde{\bar{U}}_R^b\tilde{U}_R^c\right)\left(  \tilde{\bar{U}}_L^b\tilde{U}_L^c+\tilde{\bar{D}}_L^b\tilde{D}_L^c\right.\right.\nonumber\\
&+&\left.\left.\tilde{U}_R^b\tilde{\bar{U}}_R^c\right) \right]-y_U^2\left[  \left(\tilde{H}_1\tilde{D}_L^a-\tilde{H}_2\tilde{U}_L^a\right)\left(\tilde{\bar{H}}_1\tilde{\bar{D}}_L^a-\tilde{\bar{H}}_2\tilde{\bar{U}}_L^a\right) +\tilde{U}_R^a\tilde{\bar{U}}_R^a\left(\tilde{H}_1\tilde{\bar{H}}_1+\tilde{H}_2\tilde{\bar{H}}_2\right)\right.\nonumber\\
&+&\left.\tilde{U}_R^a\tilde{\bar{U}}_R^b\left(\tilde{\bar{U}}_L^a\tilde{U}^b_L+\tilde{\bar{D}}_L^a\tilde{D}_L^b\right)  \right]-y^2_N\left[ \left( \tilde{\bar{N}}_L \tilde{\bar{H}}_2 -\tilde{\bar{E}}_L \tilde{\bar{H}}_1 \right)\left( \tilde{N}_L \tilde{H}_2 -\tilde{E}_L \tilde{H}_1 \right)\right.\nonumber\\
&+&\left.\tilde{N}_R\tilde{\bar{N}}_R\left(\tilde{H}_1\tilde{\bar{H}}_1+\tilde{H}_2\tilde{\bar{H}}_2+\tilde{N}_L\tilde{\bar{N}}_L+\tilde{E}_L\tilde{\bar{E}}_L\right) \right]
-y^2_E\left[ \left( \tilde{\bar{N}}_L \tilde{\bar{H}}^{\prime}_2 -\tilde{\bar{E}}_L \tilde{\bar{H}}^{\prime}_1 \right)\left( \tilde{N}_L \tilde{H}^{\prime}_2 -\tilde{E}_L \tilde{H}^{\prime}_1 \right)\right.\nonumber\\
&+&\left.\tilde{E}_R\tilde{\bar{E}}_R\left(\tilde{H}^{\prime}_1\tilde{\bar{H}}^{\prime}_1+\tilde{H}^{\prime}_2\tilde{\bar{H}}^{\prime}_2+\tilde{N}_L\tilde{\bar{N}}_L+\tilde{E}_L\tilde{\bar{E}}_L\right) \right]
-y_R^2\left( \tilde{U}_R^a \tilde{U}_R^a \tilde{\bar{U}}_R^b \tilde{\bar{U}}_R^b+4 \tilde{\bar{U}}_R^a \tilde{U}_R^a \tilde{\bar{E}}_R \tilde{E}_R\right) \nonumber\\
&+&\left\{ \sqrt{2}y_U g_{N4} \epsilon^{abc}\left[\tilde{U}_L^b\tilde{D}_L^c\left(\tilde{\bar{H}}_1\tilde{\bar{D}}^a_L-\tilde{\bar{H}}_2\tilde{\bar{U}}^a_L\right) \right. +\tilde{U}^a_R\tilde{\bar{U}}_R^b\left(\tilde{U}^c_L\tilde{\bar{H}}_1+\tilde{D}^c_L\tilde{\bar{H}}_2\right)\right] \nonumber\\
&-&y_U y_N \tilde{U}^a_R\tilde{\bar{N}}_R \left(\tilde{\bar{U}}^a_L\tilde{N}_L+\tilde{\bar{D}}^a_L\tilde{E}_L\right)-
y_N y_E\left.\tilde{N}_R\tilde{\bar{E}}_R\left( \tilde{\bar{H}}_1\tilde{H}^{\prime}_1+\tilde{\bar{H}}_2\tilde{H}^{\prime}_2 \right)\right. \nonumber\\
&+&y_R \tilde{\bar{U}}_R^a \left[2 \sqrt{2}g_{N4} \epsilon^{abc}\tilde{\bar{U}}_L^b \tilde{\bar{D}}_L^c \tilde{\bar{E}}_R+2 y_U \tilde{\bar{E}}_R \left( \tilde{\bar{D}}^a_L \tilde{\bar{H}}_1 -  \tilde{\bar{U}}^a_L\tilde{\bar{H}}_2 \right)
+y_E \tilde{\bar{U}}_R^a \left( \tilde{\bar{E}}_L \tilde{\bar{H}}_1^\prime- \tilde{\bar{N}}_L \tilde{\bar{H}}_2^\prime \right) \right]
\nonumber\\
&+&\left.h.c.\right\}+{\cal L}_{mix},
\label{LF4}
\ee
with ${\cal L}_{mix}$ defined in function of the $F$ auxiliary fields associated with the MSSM two Higgs super-doublets:
\be
{\cal L}_{mix}&=&-\sum_{\phi_p} \left(F_{\phi_p,N4} F^\dagger_{\phi_p,MSSM}+h.c.\right),\ \phi_p=H^\prime_1,H^\prime_2,H_1,H_1,\ 
F_{H^\prime_1,N4}=-y_E\tilde{E}_L\tilde{\bar{E}}_R,\nonumber\\ 
F_{H^\prime_2,N4}&=&y_E\tilde{N}_L\tilde{\bar{E}}_R,\ F_{H_1,N4}=-y_U\tilde{D}_L^a\tilde{\bar{U}}_R^a-y_N\tilde{E}_L\tilde{\bar{N}}_R,\ F_{H_2,N4}=y_U\tilde{U}_L^a\tilde{\bar{U}}_R^a+y_N\tilde{N}_L\tilde{\bar{N}}_R.\nonumber\\ 
\label{Lmix}
\ee
The corresponding MSSM auxiliary fields $F$ can be found in \cite{Martin:1997ns} and references therein. Also, in the Eqs.(\ref{LF4},\ref{Lmix}) we used $\tilde{H}$ and $\tilde{H}^{\prime}$ to indicate the scalar Higgs  doublets, for consistency with the rest of the notation where the tilde identifies the scalar component of a chiral superfield or the fermionic component of a vector superfield. The remaining Yukawa interaction terms are determined by the superpotential, and can be expressed as
\be
{\cal L}_{P-Yuk}&=&\sqrt{2}g_{N4}\epsilon^{abc}\left(U_L^a D_L^b\tilde{\bar{U}}_R^c+U_L^a \tilde{D}_L^b\bar{U}_R^c+\tilde{U}_L^a D_L^b\bar{U}_R^c\right) +  y_U  \left[\left( H_1 D_L^a -H_2U_L^a\right)\tilde{\bar{U}}_R^a\right.\nonumber\\
&+&\left.\left( \tilde{H}_1 D_L^a -\tilde{H}_2U_L^a\right)\bar{U}_R^a+\left( H_1 \tilde{D}_L^a -H_2\tilde{U}_L^a\right)\bar{U}_R^a\right]+y_N  \left[\left( H_1 E_L -H_2N_L\right)\tilde{\bar{N}}_R\right.     \nonumber\\
&+&\left.\left( H_1 \tilde{E}_L -H_2\tilde{N}_L\right)\bar{N}_R+\left( \tilde{H}_1 E_L -\tilde{H}_2N_L\right)\bar{N}_R\right]+y_E  \left[\left( H^{\prime}_1 E_L -H^{\prime}_2N_L\right)\tilde{\bar{E}}_R\right.  \nonumber\\
&+&\left.\left( H^{\prime}_1 \tilde{E}_L -H^{\prime}_2\tilde{N}_L\right)\bar{E}_R+\left( \tilde{H}^{\prime}_1 E_L -\tilde{H}^{\prime}_2N_L\right)\bar{E}_R\right]-y_R \bar{U}^a_R \left( \bar{U}_R^a \tilde{\bar{E}}_R+ \bar{\tilde{U}}_R^a \bar{E}_R \right)  
\nonumber\\
&+&h.c..
\ee
The soft SUSY breaking terms, finally, can be written straightforwardly starting from the superpotential in Eq.(\ref{spmwt}), to which we add the \nf gaugino and scalar mass terms as well:
\be
\label{Lsof}
{\cal L}_{soft}&=&-\left[ a_{N4}\epsilon^{abc}\tilde{U}_L^a\tilde{D}_L^b\tilde{\bar{U}}_R^c+a_U\left(\tilde{H}_1\tilde{D}_L^a - \tilde{H}_2\tilde{U}_L^a\right) \tilde{\bar{U}}^a_R+a_N\left(\tilde{H}_1\tilde{E}_L-\tilde{H}_2\tilde{N}_L\right) \tilde{\bar{N}}_R\right.\nonumber\\
&+&\left.a_E\left(\tilde{H}^{\prime}_1\tilde{E}_L-\tilde{H}^{\prime}_2\tilde{N}_L\right) \tilde{\bar{E}}_R+a_R \tilde{\bar{U}}_R^a \tilde{\bar{U}}_R^a \tilde{\bar{E}}_R + \frac{1}{2}M_D\bar{D}^a_R\bar{D}^a_R+c.c.\right]-M^2_Q\tilde{\bar{Q}}^a_L\tilde{Q}^a_L\nonumber\\
&-&M^2_U\tilde{\bar{U}}^a_R\tilde{U}^a_R - M^2_L\tilde{\bar{L}}_L\tilde{L}_L-M^2_N\tilde{\bar{N}}_R\tilde{N}_R-M^2_E\tilde{\bar{E}}_R\tilde{E}_R.
\ee

\section{Scalar Squared Mass Matrices}
\label{appMs2}
The 4SYM Higgs squared mass matrix is
\be
\label{mtch2}
{\cal M}^2_{N4\textrm{-}h}&=&\frac{1}{2}\left(
\begin{array}{cc}
 g_{N4}^2 v_{N4}^2 & -g_{N4}^2 v_{N4}^2 \\
 -g_{N4}^2 v_{N4}^2 & g_{N4}^2 v_{N4}^2
\end{array}
\right)\, ,\ \left({\cal M}^2_{N4\textrm{-}h}\right)_{ij}=\left.\frac{\partial^2 V}{\partial \phi^{N4\textrm{-}h}_i \partial \phi^{N4\textrm{-}h}_j}\right|_{\phi=\left<  \phi \right>}\, ,\nonumber\\ 
\phi^{N4\textrm{-}h}&=&\Re\left( \frac{\tilde{D}^1_L- i \tilde{D}^2_L}{\sqrt{2}},\frac{\tilde{D}^1_L+ i \tilde{D}^2_L}{\sqrt{2}} \right)\,, m_{h^{N4}}=g_{N4} v_{N4}\,.
\ee
The massless eigenstate in the last matrix is the longitudinal degree of freedom of the $N4$-photon $G$ in Eq.(\ref{g0e}):
\beq
\pi_{N4}=\frac{1}{\sqrt{2}}\left( 1,1 \right)\cdot\phi^{N4\textrm{-}h}\,.
\eeq
The charged-Higgs squared mass matrix is
\be
\label{mch2}
{\cal M}^2_{h^\pm}&=&\left(\begin{array}{cc}{\cal M}^2_{hc} & 0 \\0 & {\cal M}^2_{hl}\end{array}\right),
\ee
\be
\label{mhc2}
\left({\cal M}^2_{hc}\right)_{11}&=&  \frac{1}{4} \left(4 b \text{ct}_{\beta }+c_{\beta }^2 g_L^2 v_H^2-v_{N4}^2 \left(g_L^2-2 y_U^2\right)\right)\,,\ \left({\cal M}^2_{hc}\right)_{12}=b+\frac{1}{4} c_{\beta } g_L^2 v_H^2 s_{\beta } \nonumber\\ 
 \left({\cal M}^2_{hc}\right)_{13}&=&\frac{1}{4} v_H   s_{\beta } v_{N4} \left(g_L^2-2 y_U^2\right)\,,\ \left({\cal M}^2_{hc}\right)_{14}=-\frac{a_U v_{N4}}{\sqrt{2}}\ ,\nonumber\\
 \left({\cal M}^2_{hc}\right)_{22}&=& b t_{\beta }+\frac{1}{4} g_L^2 \left(v_H^2 s_{\beta }^2+v_{N4}^2\right)\,,\ \left({\cal M}^2_{hc}\right)_{23}= \frac{1}{4} c_{\beta } g_L^2 v_H v_{N4}\ ,\nonumber\\
 \left({\cal M}^2_{hc}\right)_{24}&=& -\frac{\mu  v_{N4} y_U}{\sqrt{2}}\,,\ \left({\cal M}^2_{hc}\right)_{33}= \frac{1}{4} v_H^2 \left(c_{2 \beta } g_L^2+2 s_{\beta }^2 y_U^2\right)\,, \nonumber\\
 \left({\cal M}^2_{hc}\right)_{34}&=& \frac{1}{\sqrt{2}}v_H \left(a_U s_{\beta }-\mu  c_{\beta } y_U\right)\,,\nonumber\\
\left({\cal M}^2_{hc}\right)_{44}&=& \frac{1}{4} \left(g_Y^2 \left(c_{2 \beta } v_H^2-v_{N4}^2\right)+2 y_U^2 \left(v_H^2 s_{\beta }^2+v_{N4}^2\right)+4 M_U^2\right),\nonumber\\
 \left({\cal M}^2_{hl}\right)_{11}&=&M_L^2+\frac{1}{2} s_{\beta }^2 v_H^2 y_N^2+\frac{1}{8}\left(g_L^2+3 g_Y^2\right) \left(c_{2 \beta } v_H^2-v_{\text{N4}}^2\right)\,,\nonumber\\ 
 \left({\cal M}^2_{hl}\right)_{12}&=&\frac{1}{\sqrt{2}}v_H \left(a_N s_{\beta }-\mu c_{\beta } y_N\right), \nonumber\\
\left({\cal M}^2_{hl}\right)_{22}&=&M_N^2+\frac{1}{4}g_Y^2 v_{\text{N4}}^2+\frac{1}{4}v_H^2 \left(y_N^2-c_{2 \beta } \left(g_Y^2+y_N^2\right)\right),
\ee
\beq
\left({\cal M}^2_{h^\pm}\right)_{ij}=\left.\frac{\partial^2 V}{\partial \phi^{h^\pm}_i \partial \phi^{h^\pm}_j}\right|_{\phi=\left<  \phi \right>}\, ,\ 
\phi^{h^\pm}=\Re\left(\tilde{H}_1, \tilde{H}^{\prime}_2, \tilde{U}^3_L, \tilde{\bar{U}}^3_R, \tilde{N}_L, \tilde{\bar{N}}_R, \right)\,.
\eeq
The massless eigenstate in the Hermitian matrix ${\cal M}^2_{hc}$, Eq(\ref{mhc2}), is the longitudinal degree of freedom of the $W$ gauge boson:
\beq
\pi_W=N_W \left( s_\beta v_{H},  -c_\beta v_{H},  v_{N4} \right)\cdot\phi^{h^\pm}\,,\ N_W^{-2}=v^2_{N4} + v^2_H\,.
\eeq
The remaining eigenvalues of  ${\cal M}^2_{hc}$ and those of ${\cal M}^2_{hl}$ are all non-zero: they have rather lengthy and not particularly instructive expressions, and therefore we do not write them here.

The $N4$-charged Higgs squared mass matrix is
\be
\label{mtcch2}
{\cal M}^2_{N4 \textrm{-}h^\pm}=\left(\begin{array}{cc}{\cal M}^2_d & -{\cal M}^2_o \\  {\cal M}^{2}_o & {\cal M}^2_d\end{array}\right),
\ee
\be
\left({\cal M}^2_{d}\right)_{11}&=& \frac{1}{4}c_{2 \beta } g_L^2 v_H^2+\frac{1}{2} s_{\beta }^2 y_U^2 v_H^2-\frac{1}{4}\left(g_L^2-4 g_{N4}^2\right) v_{N4}^2\,,\ \left({\cal M}^2_{d}\right)_{12}=\frac{1}{ \sqrt{2}} v_H \left(a_U s_{\beta }-\mu  c_{\beta } y_U\right) \nonumber\\
 \left({\cal M}^2_{d}\right)_{22}&=& M_U^2+\frac{1}{4}\left(4 g_{N4}^2-g_Y^2\right) v_{N4}^2+\frac{1}{4}v_H^2 y_U^2+\frac{1}{4}c_{2 \beta } v_H^2 \left(g_Y^2-y_U^2\right)\,,\nonumber\\
\left({\cal M}^2_{o}\right)_{ij}&=&\frac{1}{\sqrt{2}} a_{N4} v_{VC}\epsilon_{ij}\,,\ \left({\cal M}^2_{N4\textrm{-}h^\pm}\right)_{ij}=\left.\frac{\partial^2 V}{\partial \phi^{N4\textrm{-}h^\pm}_i \partial \phi^{N4\textrm{-}h^\pm}_j}\right|_{\phi=\left<  \phi \right>}\, ,\nonumber\\
\phi^{N4\textrm{-}h^\pm}&=&\Re\left( \frac{\tilde{U}^1_L- i \tilde{U}^2_L}{\sqrt{2}},\frac{\tilde{\bar{U}}^1_R+ i \tilde{\bar{U}}^2_R}{\sqrt{2}} \right)\bigcup\Im\left( \frac{\tilde{U}^1_L- i \tilde{U}^2_L}{\sqrt{2}},\frac{\tilde{\bar{U}}^1_R+ i \tilde{\bar{U}}^2_R}{\sqrt{2}} \right)\,.
\ee
The doubly charged-Higgs squared mass matrix is
\be
\left({\cal M}^2_{h^{2\pm}}\right)_{11}&=&M_L^2+\frac{1}{2}c_{\beta }^2 v_H^2 y_E^2-\frac{1}{8}\left(g_L^2-3 g_Y^2\right) \left(c_{2 \beta } v_H^2-v_{\text{N4}}^2\right),\\
 \left({\cal M}^2_{h^{2\pm}}\right)_{12}&=&\frac{1}{\sqrt{2}} v_H \left(\mu  s_{\beta } y_E-a_E c_{\beta }\right), \left({\cal M}^2_{h^{2\pm}}\right)_{22}= \frac{1}{2}\left(v_{\text{N4}}^2-\frac{1}{2}c_{2 \beta } v_H^2\right) g_Y^2+M_E^2+\frac{1}{2}c_{\beta }^2 v_H^2 y_E^2 \nonumber
\ee
\beq
\left({\cal M}^2_{h^{2\pm}}\right)_{ij}=\left.\frac{\partial^2 V}{\partial \phi^{h^{2\pm}}_i \partial \phi^{h^{2\pm}}_j}\right|_{\phi=\left<  \phi \right>}\, ,\ 
\phi^{h^{2\pm}}=\Re\left(\tilde{E}_L, \tilde{\bar{E}}_R, \right)\,.
\eeq
The eigenvalues of ${\cal M}^2_{h^{2\pm}}$ and ${\cal M}^2_{N4 \textrm{-}h^\pm}$ are all non-zero: they have rather lengthy and not particularly instructive expressions, and therefore we do not write them here.

\section{Renormalization Group Equations}
\label{rge}
In the following we write the two loop beta functions \cite{Martin:1993zk} of the gauge couplings. Notice that while the one loop beta function of $g_{N4}$ is zero the running of the coupling at two loops is non-trivial.
\beq
\frac{d g_a}{d t}=\frac{1}{16 \pi^2} \beta_a^{(1)}+\frac{1}{\left(16 \pi^2\right)^2} \beta_a^{(2)};\ g_1=g_Y,\,g_2=g_L\,,g_3=g_C\,,g_4=g_{N4};\, t=\log\left(E/m_Z\right);
\eeq
\beq
\beta_1^{(1)}=15 g_1^3,
\eeq
\beq
\beta_1^{(2)}=-\frac{42}{5} g_1^3 y_N^2-\frac{26}{5} g_1^3 y_t^2-\frac{108}{5} g_1^3 y_{\text{N4}}^2-\frac{54}{5} g_1^3 y_U^2-\frac{78}{5} g_1^3 y_E^2+\frac{1297 g_1^5}{25}+\frac{81}{5} g_2^2 g_1^3+\frac{88}{5} g_3^2 g_1^3+\frac{108}{5} g_4^2 g_1^3,
\eeq
\beq
\beta_2^{(1)}=3 g_2^3,
\eeq
\beq
\beta_2^{(2)}=-2 g_2^3 y_N^2-6 g_2^3 y_t^2-12 g_2^3 y_{\text{N4}}^2-6 g_2^3 y_U^2-2 g_2^3 y_E^2+39 g_2^5+\frac{27}{5} g_1^2 g_2^3+24 g_3^2 g_2^3+12 g_4^2 g_2^3,
\eeq
\beq
\beta_3^{(1)}=-3 g_3^3,
\eeq
\beq
\beta_3^{(2)}=-4 g_3^3 y_t^2+14 g_3^5+\frac{11}{5} g_1^2 g_3^3+9 g_2^2 g_3^3,
\eeq
\beq
\beta_{4}^{(1)}=0,
\eeq
\beq
\beta_{4}^{(2)}=-48 g_4^3 y_{\text{N4}}^2-16 g_4^3 y_U^2+48 g_4^5+\frac{36}{5} g_1^2 g_4^3+12 g_2^2 g_4^3.
\eeq
In the following we write the beta functions at two loops of the Yukawa couplings appearing in the superpotential Eq.(\ref{spmwt}) and of that of the top quark. Notice that we substituted $g_{N4}$ in the superpotential with $y_{N4}$, since their respective beta functions are indeed different, and assumed $y_R=0$, as we did in the rest of the paper. All the beta functions below are divided by the respective Yukawa coupling.
\beq
y_p^{-1}\frac{d y_p}{d t}=\frac{1}{16 \pi^2} \beta_p^{\prime(1)}+\frac{1}{\left(16 \pi^2\right)^2} \beta_p^{\prime(2)};\ p=N4,U,N,E,t\,;
\eeq
\beq
\beta_{N4}^{\prime(1)}=-\frac{9 g_1^2}{5}-3 g_2^2-12 g_4^2+12 y_{\text{N4}}^2+4 y_U^2,
\eeq
\be
\beta_{N4}^{\prime(2)}&=&\frac{36}{5} g_1^2 y_{\text{N4}}^2+12 g_2^2 y_{\text{N4}}^2+48 g_4^2 y_{\text{N4}}^2+\frac{6}{5} g_1^2 y_U^2+6 g_2^2 y_U^2+\frac{1431 g_1^4}{50}+\frac{9}{5} g_2^2 g_1^2+\frac{72}{5}
   g_4^2 g_1^2+\frac{27 g_2^4}{2}+48 g_4^4\nonumber\\
   &+&24 g_2^2 g_4^2-4 y_N^2 y_U^2-12 y_t^2 y_U^2-48 y_{\text{N4}}^2 y_U^2-96 y_{\text{N4}}^4-18 y_U^4,
\ee
\beq
\beta_U^{\prime(1)}=-\frac{9 g_1^2}{5}-3 g_2^2-8 g_4^2+y_N^2+3 y_t^2+8 y_{\text{N4}}^2+6 y_U^2,
\eeq
\be
\beta_U^{\prime(2)}&=&\frac{18}{5} g_1^2 y_N^2+\frac{4}{5} g_1^2 y_t^2+16 g_3^2 y_t^2+\frac{12}{5} g_1^2 y_{\text{N4}}^2+12 g_2^2 y_{\text{N4}}^2+32 g_4^2 y_{\text{N4}}^2+\frac{18}{5} g_1^2 y_U^2+6 g_2^2
   y_U^2+24 g_4^2 y_U^2\nonumber\\
   &+&\frac{1431 g_1^4}{50}+\frac{9}{5} g_2^2 g_1^2+12 g_4^2 g_1^2+\frac{27 g_2^4}{2}+32 g_4^4+12 g_2^2 g_4^2-3 y_N^2 y_U^2-3 y_N^4-y_E^2 y_N^2-9 y_t^2 y_U^2-9
   y_t^4\nonumber\\
   &-&56 y_{\text{N4}}^2 y_U^2-64 y_{\text{N4}}^4-22 y_U^4,
\ee
\beq
\beta_N^{\prime(1)}=-\frac{21 g_1^2}{5}-3 g_2^2+4 y_N^2+3 y_t^2+3 y_U^2+y_E^2,
\eeq
\be
\beta_N^{\prime(2)}&=&6 g_1^2 y_N^2+6 g_2^2 y_N^2+\frac{4}{5} g_1^2 y_t^2+16 g_3^2 y_t^2+\frac{18}{5} g_1^2 y_U^2+24 g_4^2 y_U^2+\frac{12}{5} g_1^2 y_E^2+\frac{3591 g_1^4}{50}+9 g_2^2 g_1^2+\frac{27
   g_2^4}{2}\nonumber\\
   &-&9 y_N^2 y_t^2-9 y_N^2 y_U^2-10 y_N^4-3 y_E^2 y_N^2-9 y_t^4-24 y_{\text{N4}}^2 y_U^2-9 y_U^4-3 y_E^4,
\ee
\beq
\beta_E^{\prime(1)}=-\frac{39 g_1^2}{5}-3 g_2^2+y_N^2+4 y_E^2,
\eeq
\beq
\beta_E^{\prime(2)}=-\frac{6}{5} g_1^2 y_N^2+6 g_1^2 y_E^2+6 g_2^2 y_E^2+\frac{7371 g_1^4}{50}+9 g_2^2 g_1^2+\frac{27 g_2^4}{2}-3 y_N^2 y_t^2-3 y_N^2 y_U^2-3 y_N^4-3 y_E^2 y_N^2-10 y_E^4,
\eeq
\beq
\beta_t^{\prime(1)}=-\frac{13 g_1^2}{15}-3 g_2^2-\frac{16 g_3^2}{3}+y_N^2+6 y_t^2+3 y_U^2,
\eeq
\be
\beta_t^{\prime(2)}&=&\frac{18}{5} g_1^2 y_N^2+\frac{6}{5} g_1^2 y_t^2+6 g_2^2 y_t^2+16 g_3^2 y_t^2+\frac{18}{5} g_1^2 y_U^2+24 g_4^2 y_U^2+\frac{6019 g_1^4}{450}+g_2^2 g_1^2+\frac{136}{45} g_3^2
   g_1^2+\frac{27 g_2^4}{2}\nonumber\\
   &-&\frac{16 g_3^4}{9}+8 g_2^2 g_3^2-3 y_N^2 y_t^2-3 y_N^4-y_E^2 y_N^2-9 y_t^2 y_U^2-22 y_t^4-24 y_{\text{N4}}^2 y_U^2-9 y_U^4.
\ee

\end{document}